\documentclass[12pt]{iopart}

\usepackage{graphicx}           
\usepackage{latexsym}
\begin{document}
\title[The numerical computer experiment for the neutrino events........    edited Feb.08,'08] 
{The numerical computer experiment for the neutrino events with the SK neutrino oscillation parameters occurring outside the Superkamiokande detector\\
---A Detailed Analysis of \textit{Upward Through Going Muon Events} and \textit{Upward Stopping Muon Events}---}
\author{N\ Takahashi, E\ Konishi and A\ Misaki$^*$}
\address{Graduate School of Science and Technology, Hirosaki University,  Hirosaki, 036-8561 Japan}
\ead{taka@cc.hirosaki-u.ac.jp}
\address{$^*$Research Institute for Science and Engineering, Waseda University, Tokyo, 169-0092 Japan}
\date{\today}
\begin{abstract}
Adopting neutrino oscillation parameters obtained by Super-Kamiokande, a numerical computer experiment for neutrino events occurring outside the detector, is carried out in the same SK live days, 1645.9 live days, constructing the virtual Super-Kamiokande detector in the computer. The numerical results by the computer experiment could be directly compared with the real SK experimental data. The comparison shows that it is difficult to obtain convincing conclusion on the existence of the neutrino oscillation with specified neutrino oscillation parameters claimed by SK through analysis for neutrino events occurring outside the detector.
\end{abstract}
\section{Introduction}
The present situation around $\nu_{\mu}-\nu_{\tau}$ oscillation initiated by Super-Kamiokande ---hereafter, SK, simply--- is to be described as follows if we follow majority opinion:\\
(1) SK found anomaly in the atmospheric neutrinos (muon deficit)~\cite{Hirata.K.1998_1992},\\
(2) SK found anomaly in the zenith angle distribution~\cite{Fukuda.Y.1994},\\
(3) SK discovered atmospheric neutrino oscillation~\cite{Kajita.T.1999_Fukuda.Y.1998},\\
(4) Soudan 2 and MACRO  confirmed SK results~\cite{Mann.WA.2001_Ambrosio.M.2000},\\
(5) K2K confirmed atmospheric neutrino results based on the baseline experiment~\cite{Ahn.M.H.2006},\\
(6) SK found oscillations themselves directly~\cite{Ashie.Y.2004},\\
(7) MINOS begins precise experiment and confirmed SK results~\cite{Michael.D.G.2006_Blake.A.2007}.\\

As one fully understand from the history on $\nu_{\mu}-\nu_{\tau}$ oscillation on atmospheric neutrinos, it is said that  the experimental results by SK play decisively important role in $\nu_{\mu }-\nu_{\tau}$ oscillation. Therefore, to reach the final conclusions on $\nu_{\mu}-\nu_{\tau}$ oscillation by SK, it is more desirable that careful examination on the SK conclusion is performed by the minority opinion.

We have been performing Computer Numerical Experiments to re-examine the atmospheric neutrino experiment results, constructing the Virtual Super-Kamiokande apparatus in the computer and analyzing the virtual neutrino events produced both inside and outside the detector. In principle, we could obtain any physical events which SK obtain and, therefore, could compare our results with SK results directly, which may lead less ambiguous discussion around the atmospheric neutrino oscillation problems.

In the discussion on the atmospheric neutrino oscillation, there are two fundamental issues to be clarified from the experimental point of view, namely, the discrimination of the neutrino events (electron events or muon events) and the differences between the directions of the incident neutrinos and those of the produced leptons.

SK treats two different types of the neutrino events as for the points of their occurrence with regard to the detector. Ones are the neutrino events occurring inside the detector, namely, \textit{Fully Contained Events} and \textit{Partially Contained Events} in the SK terminology and the others are neutrino events occurring outside the detector, namely, \textit{Upward Through Going Muon Events} and \textit{Stopping Muon Events} in their terminology.

The mutual relation between two fundamental issues mentioned above and different types of the neutrino events is as follows: in the neutrino events occurring inside the detector, one should be careful for the discrimination between muon neutrino events and electron neutrino events, because muon neutrino events produce the different Cherenkov light patterns from those of electron neutrino events inside the detector.
\footnote{SK claim they discriminate muon neutrino events from electron neutrino events almost perfectly\cite{Kasuga.S.1996}, while Anokhina and Galkin \cite{Anokhina.A_Galkin.V.2006} and Galkin et al\cite{Galkin.V.2004.2005} have different opinions.}
Furthermore, one should be very sensitive to the differences between the directions of the incident neutrinos and those of the emitted leptons ( muons or electrons), because their scattering angles could not be neglected due to their smaller energies.

On the other hand, in the neutrino events occurring outside the detector, one need not worry about the both discrimination among the neutrino events and scattering angles of the neutrino events, because these events should be identified as the muon events due to their larger effective volume and the scattering angle of the produced leptons concerned could be neglected due to their higher energies.

In the present paper, we analyze the neutrino events occurring outside the detector. The reason why we carry out the analysis of \textit{Upward Through Going Muon Events} and \textit{Stopping Muon Events} is that this analysis is more simpler and easier than that of the \textit{Fully Contained Events} and \textit{Partially Contained Events}. Our analysis of the neutrino events occurring inside the detector will be presented as a series of the papers elsewhere.

\newpage
\section{The Algorithm of the Computer Numerical Experiment}
Our fundamental approach to examine the experimental results obtained by  SK is to carry out "re-experiment" of SK. For the purpose, we have constructed the virtual SK detector in the computer and analyze virtual "experimental" events produced in the virtual detector.

The characteristics of our Computer Numerical Experiment are that we could obtain any physical results which SK obtain and we could examine SK results more concretely, comparing SK results with our results directly.

\begin{figure}[h]
\begin{minipage}[t]{0.49\linewidth}
\centering
\includegraphics[width=18pc]{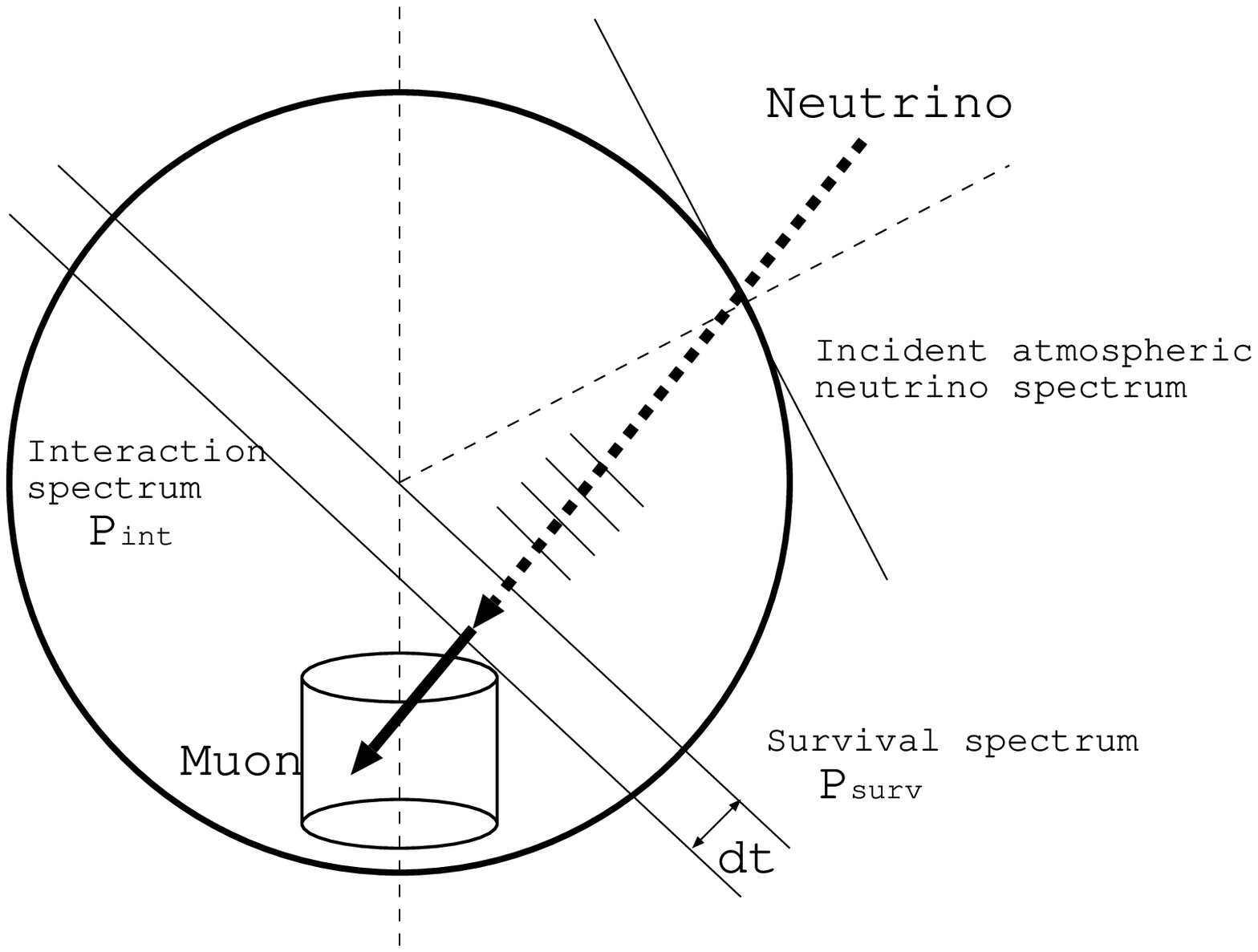}
\caption{Schematic illustration of the numerical experiment.}
\label{fig:01}
\end{minipage}\hspace{0.5pc}%
\begin{minipage}[t]{0.49\linewidth}
\centering
\includegraphics[width=18pc]{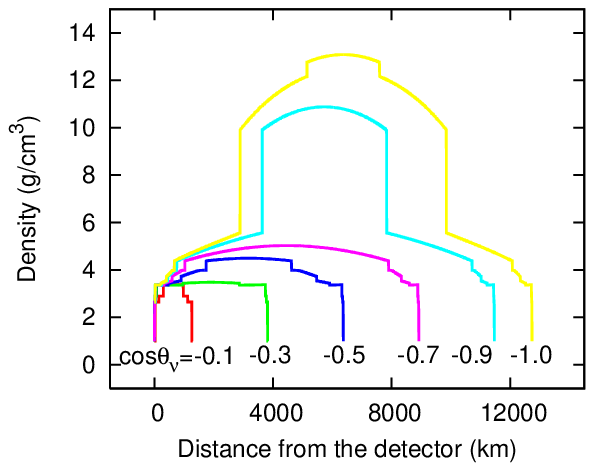}
\caption{Densities of the Earth for different zenith angles $\theta _{\nu }$.}
\label{fig:02} 
\end{minipage}
\end{figure}

Here, we give the algorithm for our computer numerical experiment. In Figure \ref{fig:01}, we show the concept of our numerical computer experiment schematically. The procedure of our numerical experiment is as follows: we start a neutrino with some energy on the opposite side to the Earth to the Super-kamiokande. For the purpose, taking the atmospheric neutrino energy spectrum on the opposite side of the Earth to the detector which SK utilizes~\cite{Honda.M.1996}, we choose a neutrino with some energy by the random sampling from this spectrum.

\vspace{-3mm}
\vspace{-3mm}
\noindent
Neutrinos thus chosen penetrate into the interior of the Earth and interact at some point inside the Earth after the traverse through the layer with different densities shown in Figure \ref{fig:02}~\cite{Dziewonski.A.1981}. The interaction points of the neutrino events concerned are determined by the random sampling, following the probability function for the given neutrino interaction mean free path where the change of the density of the Earth is taken into account, referring to the Preliminary Reference Earth model~\cite{Dziewonski.A.1981}. The interaction mean free paths of the neutrinos for given energies concerned depend on the both density of the location and the zenith angle of the incident neutrino.

\begin{figure}[h]
\begin{minipage}[t]{18pc}
\centering
\includegraphics[width=18pc]{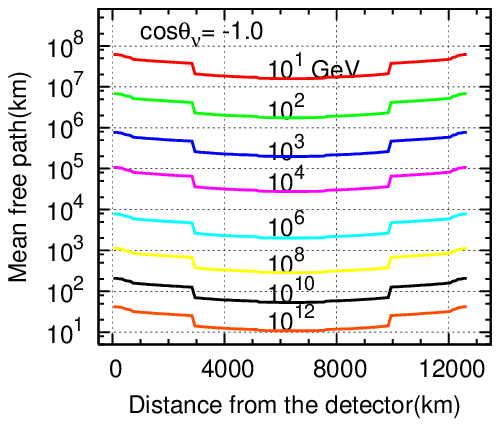}
\caption{Mean free paths of the neutrino as the function of energy for zenith angle ($\cos\theta = -1.0$) for different energies.}
\label{fig:03}
\end{minipage}\hspace{1pc}%
\begin{minipage}[t]{18pc}
\centering
\includegraphics[width=18pc]{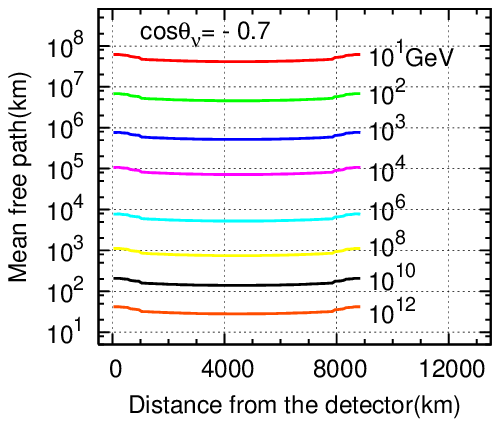}
\caption{Mean free paths of the neutrino as the function of energy for zenith angle($\cos\theta = -0.7$) for different energies.}
\label{fig:04}
\end{minipage} 
\end{figure}

In Figure \ref{fig:03} and Figure \ref{fig:04}, we give them in $\cos\theta_{\nu } = -1.0$ and $\cos\theta_{\nu } = -0.7$, where  $\cos\theta_{\nu }$ denotes zenith angle of the incident neutrino, as examples. Furthermore, the neutrino concerned produces a daughter muon by deep inelastic scattering~\cite{Kajita.T.1999_Fukuda.Y.1998}, the energy of which is also determined by the random sampling, following the probability function for the charged current (deep inelastic scattering) differential cross section concerned~\cite{Gandhi.R.H.1996_1998}.

As for \textit{Upward Going Muon Events} and \textit{Upward Stopping Muon Events} in our computer numerical experiment, we could get the essential information around them, namely, what nature the neutrinos concerned have ( normal or anti-), their location where they are produced and what energies they have, what category they belong to ( \textit{Upward Through Going Muon Events} or \textit{Upward Stopping Muon Events}  ), while the real experiments by SK give only the discrimination between  \textit{Upward Through Going Muon Events} and \textit{Upward Stopping Muon Events}.

\vspace{-3mm}
\vspace{-3mm}

Exactly speaking, the procedure which we really utilize is slightly different from what we just mention above, although it is mathematically equivalent to what we mention above.

Our real procedure adopted is as follows: instead of sampling the neutrino concerned directly from the incident neutrino energy spectrum on the surface of the Earth and pursing it from its interaction point in stochastic manner, we define the neutrino interaction probability function at $t_{n}$ which is measured from the surface of the Earth, in the following.
\vspace{-1mm}
\begin{equation}
\hspace{-25mm}
P_{int}(E_{\nu (\overline{\nu})},t_{n},\cos\theta_{\nu })dt = 
\prod _{i=1}^{n-1}
\left(1-\frac{dt}{\lambda_{i}(E_{\nu (\overline{\nu})},t_{i},\rho(\theta_{\nu },t_{i}))}\right)
\times \left(\frac{dt}{\lambda_{n}(E_{\nu(\overline{\nu})},t_{n},\rho(\theta_{\nu },t_{i}))}\right)
\label{eq:Pintdt}
\end{equation}
\vspace{1mm}
\noindent
, where $\lambda_{i}(E_{\nu (\overline{\nu})},t_{i},\rho(\theta_{\nu },t_{i}))$ 
is the mean free path of the neutrino interaction for normal and anti-neutrino with
 $E_{\nu (\overline{\nu})}$  at the depth $t_{i}$, respectively where the density of the Earth is $\rho(\theta_{\nu },t_{i})$ and $\cos\theta_{\nu }$ denotes nadir angle. Here, we consider the deep inelastic charged current interaction~\cite{Gandhi.R.H.1996_1998} and density of the Earth is taken from the Preliminary Reference Earth Model~\cite{Dziewonski.A.1981}. Here, we adopt the deep inelastic scattering exclusively as the cause of neutrino interaction, because other neutrino interactions could be neglected for our case.

By combining Eq.(\ref{eq:Pintdt}) with the atmospheric neutrino energy spectrum at the opposite side of the Earth to SK, we construct 
$N_{int, null}(E_{\nu (\overline{\nu})},t_{i},\cos\theta_{\nu })dl$, 
neutrino interaction energy spectrum without neutrino oscillations for normal and anti-neutrino which denotes frequency of the neutrino interaction in  $(t_{n}, dl)$ in the following,
\vspace{1mm}
\begin{equation}
\hspace{-15mm}
N_{int,null}(E_{\nu (\overline{\nu})},t_{n},\cos\theta_{\nu })dl =
N_{sp}(E_{\nu (\overline{\nu})},\cos\theta_{\nu })dl 
\times P_{int}(E_{\nu (\overline{\nu})},t_{n},\cos\theta_{\nu })
\label{eq:Nintnulldt}
\end{equation}
\vspace{1mm}
\noindent
, where  
$N_{sp}(E_{\nu (\overline{\nu})},\cos\theta_{\nu })$ 
denotes the atmospheric neutrino spectrum for normal and anti-neutrino at the opposite side of the Earth to SK detector~\cite{Honda.M.1996}.

The neutrino interaction energy spectrum, 
$N_{int, null}(E_{\nu (\overline{\nu})},t_{n},\cos\theta_{\nu })dl$, 
, denotes frequency for neutrino interaction in $(t_{n}, dl)$for null oscillation. The $dl$ chosen in relation to the maximum energy of neutrino energy spectrum in the following.

The maximum energies of the emitted muons are equal to energies of the parent neutrino with maximum primary energy. We utilize the incident neutrino energy spectrum by Honda et al.~\cite{Honda.M.1996} which is also utilized by SK, the maximum energy of which is $10$ TeV. Therefore, we adopt $10$ TeV as the maximum energy of muon to be pursued in our computer numerical experiment.

\vspace{-3mm}
\begin{figure}[h]
\begin{minipage}[t]{18pc}
\centering
\includegraphics[width=14pc]{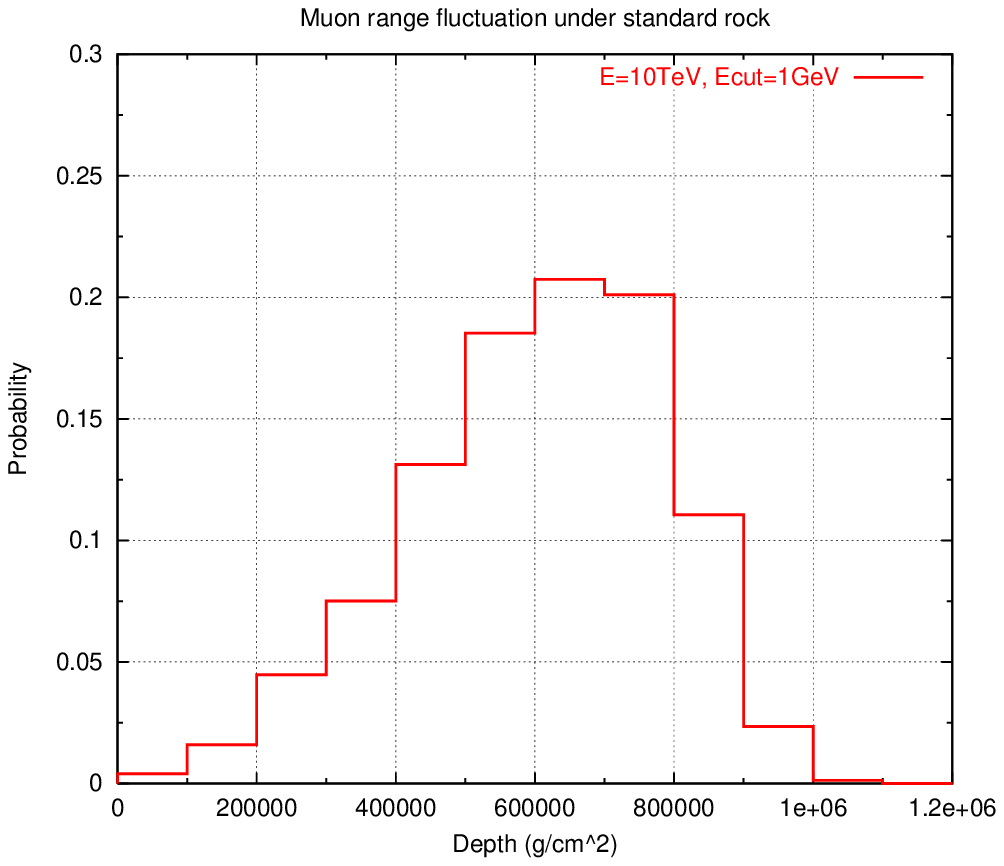}
\caption{Range Straggling of $10$ TeV muon in the standard rock.}
\label{fig:05}
\end{minipage}\hspace{1pc}%
\begin{minipage}[t]{18pc}
\centering
\includegraphics[width=14pc]{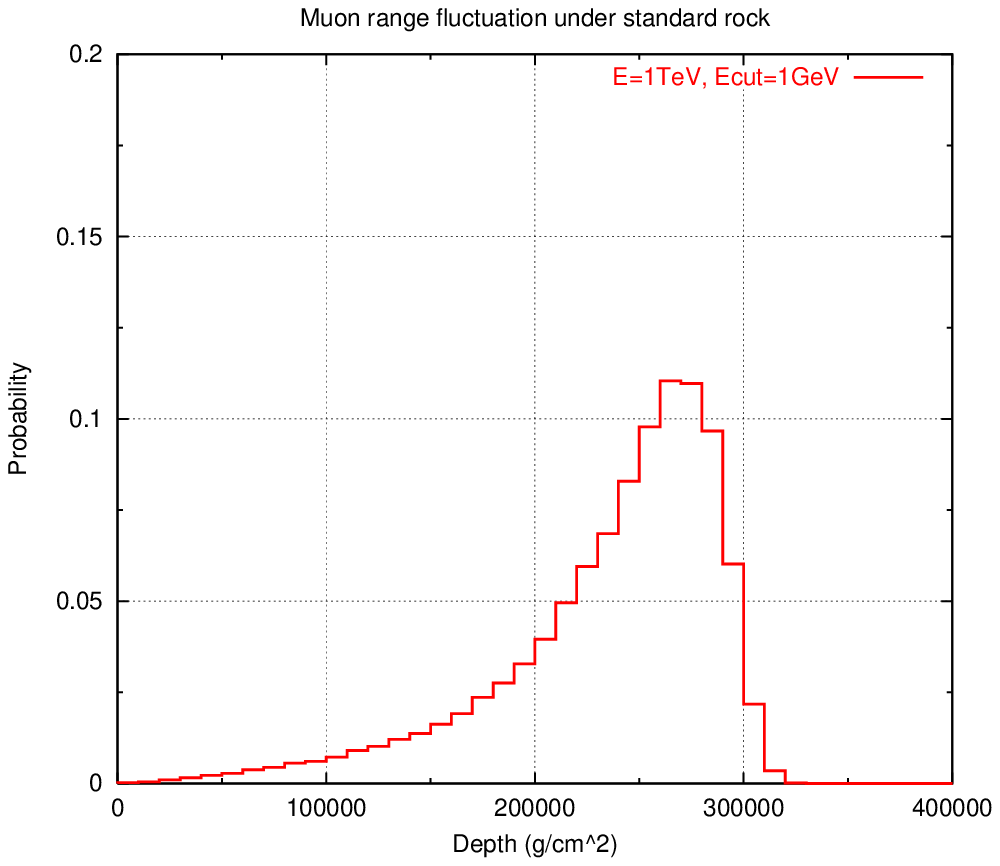}
\caption{Range Straggling of $1$ TeV muon in the standard rock.}
\label{fig:06}
\end{minipage}
\end{figure}%
\vspace{-3mm}
\noindent
Now, we decide the maximum range for $10$ TeV muon which is necessary for our computer numerical experiment. For the purpose, we simulate muons with 10 TeV in stochastic manner exactly, taking into consideration of the all processes of bremsstrahlung, direct electron pair production, nuclear interaction and ionization loss.

In Figure \ref{fig:05}, we give the range fluctuation of $10$ TeV muon thus obtained in the standard rock.  It is easily understood from the figure that the range of the muon is widely distributed by the fluctuation due to nature in these elementary processes. The shorter ranges than their average are caused by catastrophic energy loss due to the bremsstrahlung and nuclear interaction (see, for example, Figure \ref{fig:09}), while longer ranges than their average are caused essentially by lager number of the direct electron pair production ( continuous energy loss-like)(see, for example Figure \ref{fig:10}).  From Figure  \ref{fig:05}, we could adopt $2.0\times 10^{6}$ $g/cm^{2}$ as the maximum range of 10 TeV muon. In other word,  when muons with 10 TeV are produced at the place far $2.0\times 10^{6}$ $g/cm^{2}$ from the detector, such muons  never reach the detector. Therefore, it is enough for us to have interest in high energy neutrinos only which are produced within $2.0\times 10^{6}$ $g/cm^{2}$ from the detector. Now, we take $2.0\times 10^{6}$ $g/cm^{2}$ as $dl$ which is defined in Eq.(\ref{eq:Nintnulldt}). The generation points of such muons distribute uniformly within $dl$, because $dl$ is smaller by higher order magnitude of the mean free path of the neutrino for deep inelastic scattering. In Figure \ref{fig:06}, we give the range straggling with 1 TeV muon obtained in similar manner as in Figure \ref{fig:05}. From Figures \ref{fig:05} and \ref{fig:06}, it is easily supposed that the effect of the range fluctuation in higher energy muon could not be neglected in spite of the rapid decrease of the higher energy muon in muon energy spectrum\footnote{The range energy relation of high energy muon will be studied in detail in a separate paper elsewhere.} .

Our computer numerical experiment for definite zenith angle of the incident neutrino is being carried out as follow:

\vspace{3mm}
\noindent
\underline{Procedure A:} 
We calculate the neutrino interaction probability from 
Eq.(\ref{eq:Pintdt}). 
The value of $t_{n}$  is defined as the distance from the surface of the Earth to $dl$. Next, we calculate the interaction neutrino energy spectrum from 
Eq.(\ref{eq:Nintnulldt}).

\noindent
\underline{Procedure B:} 
By using 
$N_{int}(E_{\nu (\overline{\nu})},t_{n},\cos\theta_{\nu (\overline{\nu})})dl$ 
thus obtained from Eq.(\ref{eq:Nintnulldt}),
we sample randomly 
$E_{\nu (\overline{\nu})}$, 
the energy of the incident (anti-) neutrino to be pursued in the following way.
\vspace{-1mm}
\begin{equation}
\hspace{-2.5cm}
\xi=
\int^{E_{\nu  (\overline{\nu})   }}_{E_{\nu (\overline{\nu}),min}}
N_{int, null}(E_{\nu (\overline{\nu})},t_{n},\cos\theta_{\nu (\overline{\nu})})dE_{\nu (\overline{\nu})}
\Bigg/
\int^{E_{\nu (\overline{\nu}),max}}_{E_{\nu (\overline{\nu}),min}}
N_{int, null}(E_{\nu (\overline{\nu})},t_{n},\cos\theta_{\nu (\overline{\nu})})dE_{\nu (\overline{\nu})}
\label{eq:xi_Nintnull}
\end{equation}
\vspace{1mm}
\noindent
, where $E_{\nu (\overline{\nu}),max}$, $E_{\nu (\overline{\nu}),min}$ 
denote the maximum and minimum energy of the neutrino concerned, respectively. And $\xi$ denote uniform random number between (0,1).

\noindent
\underline{Procedure C:}
We determine the interaction point of the (anti-) neutrino event with energy determined by Procedure B.  By using $\xi$, the uniform random number  between (0,1), we  determine  the interaction point $\xi dl$. Namely, the neutrino interaction occurs from the detector, $t_{i}$ is given as,
\begin{equation}
t_{i}=(1-\xi)dl
\label{eq:ti}
\end{equation}
 The detailed study on the pursuit of high energy muon, relating to the range energy relation, will appear in a separate paper elsewhere.

\noindent
\underline{Procedure D:}
The $E_{\mu (\overline{\mu} )}$, the energy of the emitted (anti-)muon which occur at $(1-\xi) dl$ from the detector produced by $E_{\nu }$ is determined from the following equation,
\vspace{0mm}
\begin{equation}
\hspace{-1.5cm}
\xi_{d} = 
\int_{E_{\mu (\overline{\mu}), min}}^{E_{\mu (\overline{\mu})}}
D(E_{\nu (\overline{\nu})},E_{\mu (\overline{\mu})}) dE_{\mu (\overline{\mu})}
\Bigg/
\int_{E_{\mu (\overline{\mu}), min}}^{E_{\mu (\overline{\mu}), max}}
D(E_{\nu (\overline{\nu})},E_{\mu (\overline{\mu})}) dE_{\mu (\overline{\mu})}
\label{eq:xi_d}
\end{equation}
\vspace{1mm}
\noindent
, where $ D(E_{\nu (\overline{\nu})}, E_{\mu (\overline{\mu})}) dE_{\mu (\overline{\mu})} $ is the differential cross section of the deep inelastic scattering for (anti-) neutrino~\cite{Gandhi.R.H.1996_1998} and $E_{\mu (\overline{\mu}), max}$ equal to the $E_{\nu (\overline{\nu})}$.
The $\xi_{d}$ is a uniform random number between (0,1). 

\noindent
\underline{Procedure E:} 
For the (anti-)muon whose energy and production point is determined from Procedures C and D, we examine the behavior of the trajectory of the (anti-)muon toward the SK detector in a stochastic manner. Namely, each individual muon is pursued exactly by taking into consideration of the processes of bremsstrahlung, direct electron pair production, nuclear interaction and ionization loss in the stochastic manner without utilizing the average behavior of the muon concerned, the procedure of which is adopted by SK~\cite{Oyama.Y.1989_HaSaNi.PhDthesis}. As the result of it, we determine finally, which category each individual muon pursued  falls into: [a] stopping before it reaches the detector, [b] stopping inside detector, or [c] passing through the detector.

\vspace{3mm}
\noindent
We repeat Procedures~A to E and obtain the neutrino events without oscillation for a given live time for the real experiment.  In our computer numerical experiment, we have accumulated the events concerned which correspond to the real live time for SK. For each neutrino event, we know $E_{\nu (\overline{\nu})}$, 
the energy of the parent neutrino, $E_{\mu (\overline{\mu})}$, 
the energy of the daughter (anti-)muon,$(1-\xi)dl$, the interaction point, 
$\cos\theta_{\nu (\overline{\nu})}$ ($\cos\theta_{\mu (\overline{\mu})}$), 
the direction of both the incident (anti-)neutrino and the emitted (anti-)muon, respectively, and $t_{n}$, the distance between the interaction point of the neutrino events and the opposite side of the Earth.

\begin{figure}[t]
\centering
\includegraphics[width=20pc]{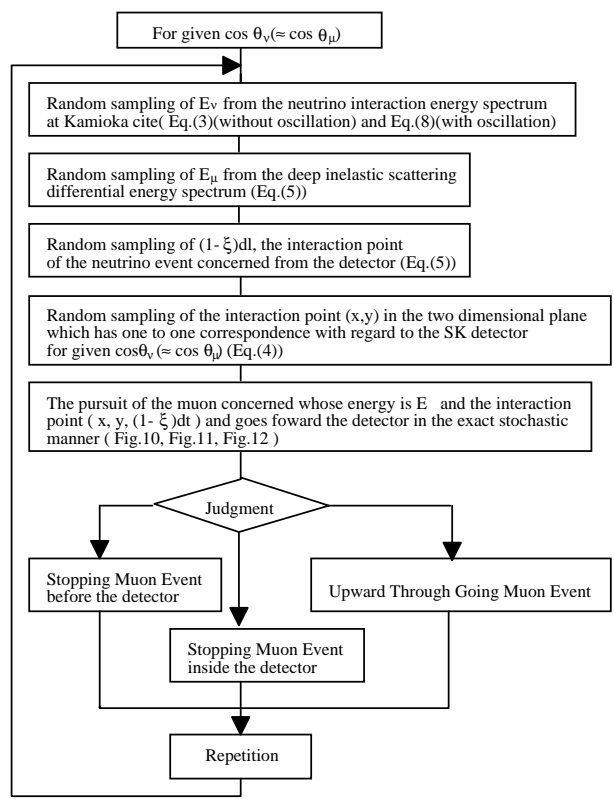}
\caption{Grand structure for the computer numerical experiment}
\label{fig:chart}
\end{figure}%

In the case without neutrino oscillation, we repeat from Procedures A to E, starting from the interaction energy spectrum(Eq.(\ref{eq:Nintnulldt})).
Finally, we obtain frequencies for \textit{Upward Stopping Muon Events} and \textit{Upward Through Going Muon Events} for given zenith angle for null oscillation. Thus, we obtain the zenith angle distributions for these categorically different events without neutrino oscillation after summation over zenith angle.

In the case with neutrino oscillation, 
$N_{int}(E_{\nu (\overline{\nu})},t_{n},\cos\theta_{\nu (\overline{\nu})})dt$ 
in 
Eq.(\ref{eq:Nintnulldt}) 
should be replaced by

\vspace{-1mm}
\begin{equation}
\hspace{-25mm}
N_{int, osc}(E_{\nu (\overline{\nu})},t_{n},\cos\theta_{\nu (\overline{\nu})})dt = 
N_{int,null}(E_{\nu (\overline{\nu})},t_{n},\cos\theta_{\nu (\overline{\nu})})dt\times 
P(\nu_{\mu} (\overline{\nu}_{\overline{\mu}}) 
\longrightarrow 
\nu_{\mu} (\overline{\nu}_{\overline{\mu}}))
\label{eq:Nintoscdt}
\end{equation}
\vspace{-1mm}
\noindent
, where 
\vspace{-3mm}
\begin{equation}
P(\nu_{\mu} (\overline{\nu}_{\overline{\mu}}) 
\longrightarrow 
\nu_{\mu} (\overline{\nu}_{\overline{\mu}}))
=
1-\sin^{2}2\theta \sin^{2}\left( \frac{1.27\Delta m^{2}(eV^{2}) t_{n}(km)}{E_{\nu (\overline{\nu})}(GeV)} \right)
\label{eq:Pnunu}
\end{equation}
\vspace{-1mm}
\noindent
, the survival probability of a given $\nu_{\mu}$.

Here, we adopt 
$\sin^{2}2\theta=1.00$ 
and 
$\Delta m^{2}=2.1 \times 10^{-3} $eV$^{2}$~\cite{Ashie.Y.2005}
, 
as obtained from SK experiment.  Other procedures are exactly same as in the case without neutrino oscillation. 

Thus, Eq.(\ref{eq:xi_Nintnull}) should be replaced by
\vspace{0mm}
\begin{equation}
\hspace{-2.3cm}
\xi=
\int^{E_{\nu (\overline{\nu})    }}_{E_{\nu (\overline{\nu}),min}}
N_{int, osc}(E_{\nu (\overline{\nu})},t_{n},\cos\theta_{\nu (\overline{\nu})})dE_{\nu (\overline{\nu})}
\Bigg/
\int^{E_{\nu (\overline{\nu}),max}}_{E_{\nu (\overline{\nu}),min}}
N_{int, osc}(E_{\nu (\overline{\nu})},t_{n},\cos\theta_{\nu (\overline{\nu})})dE_{\nu (\overline{\nu})}
\label{eq:xi_Nintosc}
\end{equation}
\vspace{0mm}
\noindent
, by using Eq.(\ref{eq:Nintoscdt}) and other procedure are not changed.

The Procedures A to E in both cases without oscillation and with oscillation is illustrated in Figure \ref{fig:chart}. 
 
\section{Results and discussion}
\subsection{The differential neutrino interaction energy spectrum with neutrino oscillation}

\begin{figure}[h]
\begin{minipage}[t]{18pc}
\centering
\includegraphics[width=18pc]{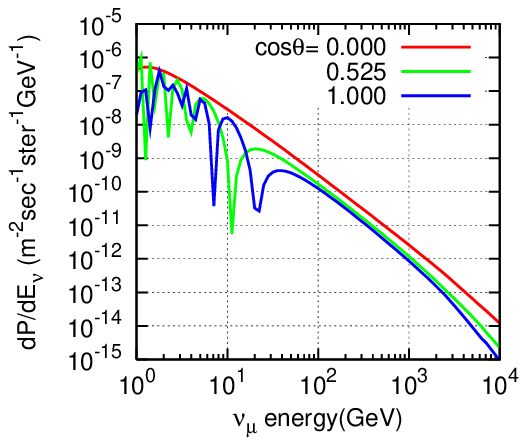}
\caption{The incident neutrino spectrum with oscillation to be sampled.}
\label{fig:07}
\end{minipage}\hspace{1pc}%
\begin{minipage}[t]{18pc}
\centering
\includegraphics[width=18pc]{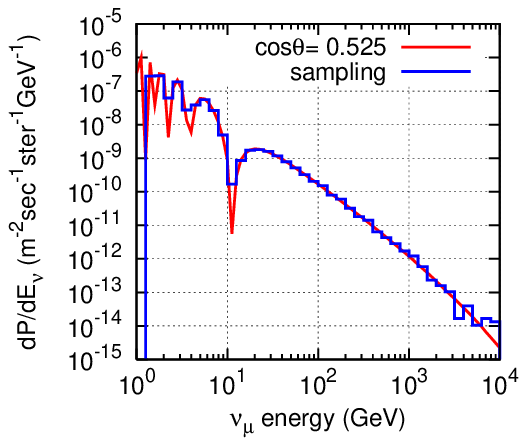}
\caption{The reproduction of the atmospheric interaction neutrino spectrum with oscillation.}
\label{fig:08}
\end{minipage}
\end{figure}

In Figure \ref{fig:07}, we give the differential neutrino interaction energy spectrum with oscillation (Eq.(\ref{eq:Nintoscdt})) for different zenith angle from which we randomly sample the energy of the neutrino event concerned.

From the figure, it is easily understood that the differential neutrino interaction energy spectrum are not so influenced by the characteristics of the nature of oscillatory above $\sim 40$ GeV under SK neutrino oscillation parameters. The result of the random sampling by Eq.(\ref{eq:xi_Nintosc}) is given in  Figure \ref{fig:08}. The excellent agreement between the theoretical curve (solid line) and the results sampled (histogram) guarantees the validity of the sampling procedure by Eq.(\ref{eq:xi_Nintosc}).

The nature of oscillatory in the neutrino interaction energy spectrum a play crucial role below $\sim 40$ GeV under the SK neutrino oscillation parameters, on which we will examine in a series of the subsequent papers in detail.

\vspace{-3mm}
\vspace{-3mm}

\subsection{Discrimination between \textit{Upward Through Going Muon Events} and \textit{Stopping Muon Events}}

The physical origin of \textit{Upward Through Going Muon Events} is the same that of \textit{Stopping Muon Events} and the only difference between them lies in the situation that the former may eventually pass through the detector, while the latter may eventually stop the both initiated by the deep inelastic scattering.

\begin{figure}[h]
\centering
\includegraphics[width=20pc]{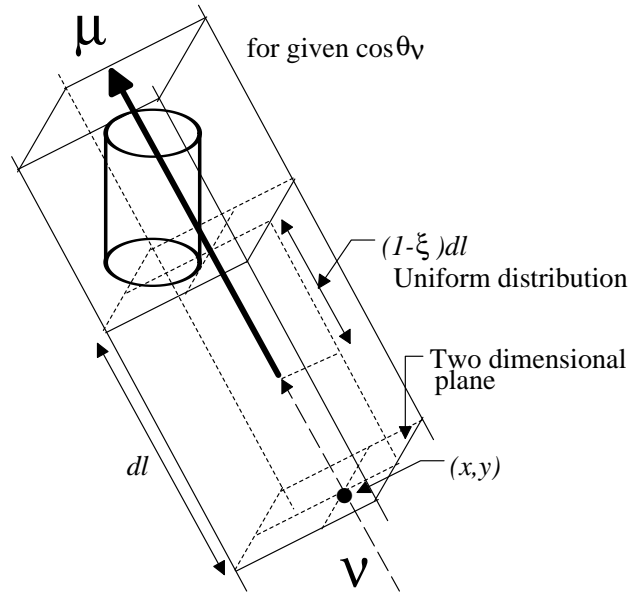}
\caption{The distribution of the generation point of the neutrino event}
\label{fig:detector_upward}
\end{figure}

The generation point of the neutrino event concerned is given Eq.(\ref{eq:ti}). As the generation point of the neutrino event distribute over  \textit{dl} ($ 2.0 \times 10^{16}$ g/cm$^{2}$ in our case ) randomly shown in Figure \ref{fig:detector_upward}, the most biggest fluctuation in the muon behavior essentially comes from the generation point of the neutrino events over $dl$. Such fluctuation could not be considered in the Detector Simulation by SK.

The energy of the neutrino is sampled randomly from Eq.(\ref{eq:xi_Nintosc}) 
and the energy of the emitted muon from the neutrino event concerned is sampled randomly from Eq.(\ref{eq:xi_d}). 
The muons whose energies and whose generation points 
thus determined are simulated in the stochastic manner exactly, taking into account of the all processes of the direct electron pair production, bremsstrahlung, nuclear interaction and ionization loss. Namely, these muons thus sampled lose their energies in stochastic manner in the course of their passages toward SK detector.

In summarized speaking, the fluctuation effect in the muon behavior secondarily comes from the range fluctuation of the muon, although this effect is rather small compared with the effect from the generation point of the neutrino events but could not be neglected.

Even if the physical natures of the muons are exactly same, some muon may be classified as the \textit{Upward Through Going Muon Events} or may be done as \textit{Stopping Muon Events} eventually due to the complicated geometry --- configuration of $(x, y)$ --- for muons concerned, which is schematically shown in Figure \ref{fig:detector_upward}. 
\vspace{0mm}

We take into account of the complicated circumstances for the muons concerned as exactly as possible. As the results of it, we could discriminate finally \textit{Upward Through Going Muon Events} from the \textit{Stopping Muon Events}.

\begin{figure}[h]
\begin{minipage}[t]{18pc}
\centering
\includegraphics[width=18pc]{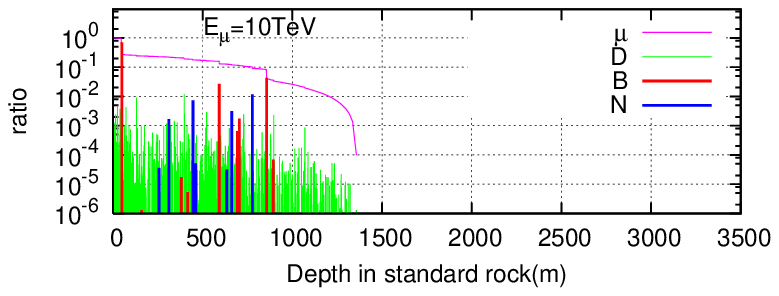}
\caption{An example of a real high energy muon behavior.}
\label{fig:09}
\end{minipage}\hspace{1pc}%
\begin{minipage}[t]{18pc}
\centering
\includegraphics[width=18pc]{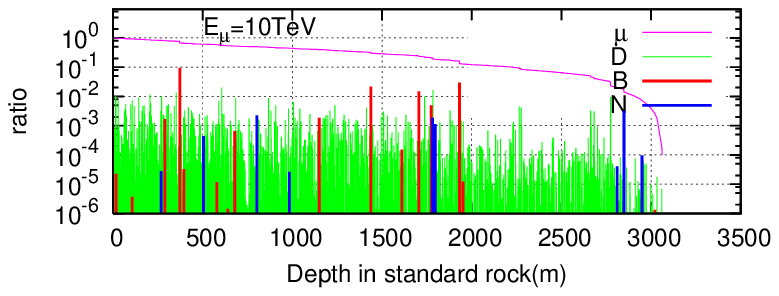}
\caption{An another example of a real high energy muon behavior.}
\label{fig:10}
\end{minipage}
\end{figure}

Among elementary processes contributed to muon energy losses, the process of the direct electron pair production, bremsstrahlung and nuclear interaction play important roles from the point of fluctuation. In Figures \ref{fig:09} and \ref{fig:10}, how these three elementary processes take place randomly in the stochastic ways are indicated as the functions of the traversed distance in standard rock by a muon with 10 TeV.   Therefore, we treat the muon energy loss in stochastic manner exactly.

\vspace{0mm}
\vspace{0mm}
In the figures, we show graphically where, which interaction (electron direct pair production, bremsstrahlung or nuclear interaction) occurs and what energy with respect to the primary energy is lost by the interaction concerned. The $\mu $ denotes the sustained energy of the muon concerned as the result of all possible energy losses. The \textbf{D} denote the energy loss due to direct electron pair production, \textbf{B} is the energy loss due to the bremsstrahlung and \textbf{N} denotes nuclear interaction, respectively. 
The Ratio denotes the ratio of energy loss concerned to the primary energy.

In Figure \ref{fig:09}, we give an example of a muon with shorter range, say, $1500$ meter.  We could understand the muon loses about $30$ \% catastrophically at the early stage by the bremsstrahlung and bigger energy losses may come from the bremsstrahlung and nuclear interaction and smaller energy losses essentially come from the direct electron pair production. In Figure \ref{fig:10} which is contrast to Figure \ref{fig:09}, we give an example of a muon with range over $3000$ meters. In this case, bremsstrahlung and nuclear interaction do not contribute to bigger energy losses and the accumulation of large number of the direct electron pair production events with rather smaller energies result in bigger energy loss.

On the contrary to our exact treatment on the individual muon behavior, SK seem to treat them in the sense of as the average theory \cite{Oyama.Y.1989_HaSaNi.PhDthesis,Ashie.Y.2005}. SK utilize the following conventional formula, revised by Lohmann, Knop and Voss~\cite{Lohman.W.1985}.
\vspace{-1mm}
\begin{equation}
\frac{dE}{dx}=a+b\times E
\label{eq:dEdx}
\end{equation}
\vspace{-1mm}
\noindent
Then, this description expresses the average energy loss of the muon which does not give the range straggling of the muon and dose not expresses the real energy loss by individual muon. When the energy of the muon is not so pretty high where we could neglect the fluctuation effect of the muon range in individual case, we could apply this expression to the analysis of real behavior of the muon. However, the energy region for muon where we are interested in is now rather high and, therefore, we should pursue the muon behavior in the stochastic manner exactly, but in not average theory, even if the fluctuation effect is taken into account \footnote{Recently, SK introduce the concept of showering muon in their treatment on the high energy upward muon event~\cite{Desai.S.2007}. In their paper, they seem to treat high energy muon in stochastic manner, utilizing GEANT3.}. It should be emphasized that the average picture of the muons obtained from the average theory like Eq.(\ref{eq:dEdx}) does not describe the individual behavior of the muon concerned and the average picture of the muons is the average behavior of the individual muon which is deviated from their average behavior when we could not neglect the fluctuation effect coming from their stochastic nature in the elementary processes.  Furthermore, it should be noticed that the stochastic treatment to the individual muon becomes important in the analysis of statistically insufficient number of the events concerned.

In summarized speaking, the fluctuation effect in the muon behavior secondarily comes from the range fluctuation of the muon, although this effect is rather small compared with the effect from the generation point of the neutrino events but could not be neglected.

\newpage
\subsection{Zenith Angle distribution for \textit{Upward Through Going Muon Events}}

\begin{figure}[h]
\begin{minipage}[t]{18pc}
\centering
\includegraphics[width=18pc]{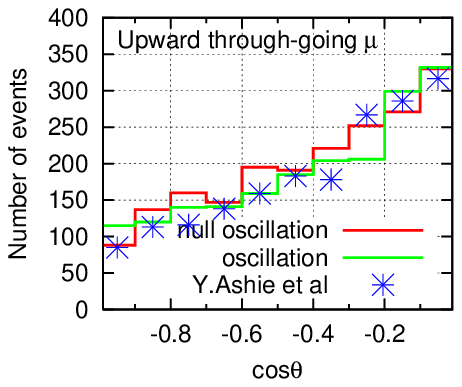}
\caption{Zenith angle distribution for the \textit{Upward Through Going Muon Events} for 1645.9 live-days.}
\label{fig:11}
\end{minipage} \hspace{1pc}%
\begin{minipage}[t]{18pc}
\centering
\includegraphics[width=18pc]{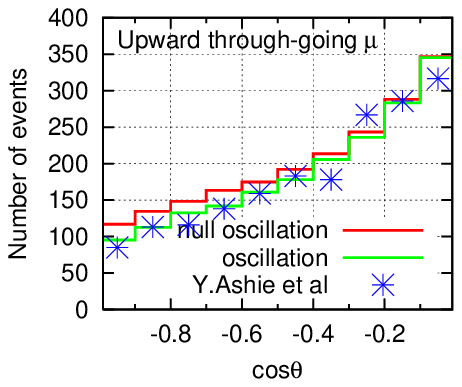}
\caption{Zenith angle distribution for the \textit{Upward Through Going Muon Events} for 164590 live-days, which is normalized to 1645.9 live-days.}
\label{fig:12}
\end{minipage}
\end{figure}

In Figure \ref{fig:11} we compare the experimental results by SK with corresponding our computer numerical results for both with neutrino oscillation and without neutrino oscillation in the case of \textit{Upward Through Going Muon Events} for the same 1645.9 live days~\cite{Ashie.Y.2005}. From the figure, we could not conclude the experimental data agree with null oscillation or with oscillation.

The 1645.9 live days is not enough for extracting definite conclusion on the neutrino oscillation under SK neutrino oscillation even if it is possible to get SK neutrino oscillation parameters through the analysis of \textit{Upward Through Going Muon Events}, because the statistical fluctuation from the average values is not small. If we compare our results with oscillation to our results without oscillation in the figure, we could find the event number with oscillation is larger than that without oscillation in the three bins  (0.0 $\sim$ -0.1),  (-0.1 $\sim$  -0.2)  and  (-0.9 $\sim$ -1.0) . Such phenomena come from the statistical fluctuation due to insufficient sampling (accumulation) of the neutrino events. 

\vspace{-3mm}
\vspace{-3mm}
\noindent
In Figure \ref{fig:12}, we give corresponding results from our computer numerical experiment for 164590 days, one hundred times of the SK real live days. Now, we understand from the figure that our results with oscillation are always smaller than that without oscillation which attributes to larger statistics. This denotes that sampling numbers for 164590 live days is quite enough for extracting the definite conclusion on the neutrino oscillation around the SK neutrino oscillation parameters.

Furthermore, it is reasonable to say from the figure that we could not discriminate the difference in both with null oscillation and with oscillation under SK oscillation parameters from our computer numerical experiment.  In other words, this is quite understandable, because the definition of the \textit{Upward through Going Muon Events} is the muon events, the minimum energy of which is only given $\sim$ 1.6 GeV and nothing more both the incident energy, say, nobody knows both generation points and termination points and energies of the events concerned, but knows their zenith angle merely. It should be rather curious, if we really could extract definite neutrino oscillation parameters from such extremely poor information around the events concerned.

In Figure \ref{fig:12}, we add, on purpose, the SK experimental data for 1645.9 live days together with our data for 164590 live days ---450 live years--- as one hundred times as SK real live days. From the comparison between SK results for 1645.9 live days and our result for 164590 live days ---450 live years---, we could not conclude directly that SK experimental data rather agree with our results with oscillation, because statistics of both results is quite different from each other and we do not allow to compare them directly\footnote{SK discusses the evidence of the neutrino oscillation under SK oscillation parameters, based on the comparison of the real experimental data (for example, for 1645 live days) with Monte Carlo Simulation data without oscillation (for example, 100 live years which are much larger than that of real experiment).  However, it seems not suitable to compare directly the one with smaller statistics with the other with further large statistics, because we have the possibility to kill real fluctuation.  SK is desirable to compare the real experimental data for 1645.9 live days with Monte Carlo simulation for the same live days to extract definite conclusion. Further more, it should be noticed that a 1645.9 live days might be different from another 1645.9 days due to significant insufficient statistics.} .

Comparing the smoothness of the histograms in Figure \ref{fig:11} with these in Figure \ref{fig:12}, we could conclude that the average values of zenith angle distributions attain sufficiently at the 'true' ones. In other words, we think SK live days does not give enough statistics to extract clear cut conclusion under SK neutrino oscillation parameters from the analysis of \textit{Upward Through Going Muon Events}, even if SK finally could give definite neutrino oscillation parameters from the analysis of \textit{Fully Contained Events} and \textit{Partially Contained Events}. 

\begin{figure}[h]
\begin{minipage}[t]{12pc}
\centering
\includegraphics[width=12pc]{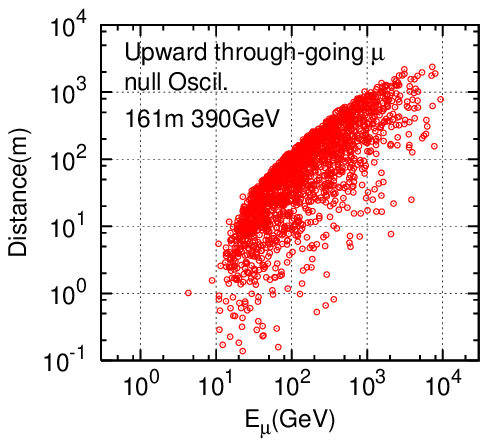}
\caption{}
\label{fig:13}
\end{minipage}\hspace{0.5pc}%
\begin{minipage}[t]{12pc}
\centering
\includegraphics[width=12pc]{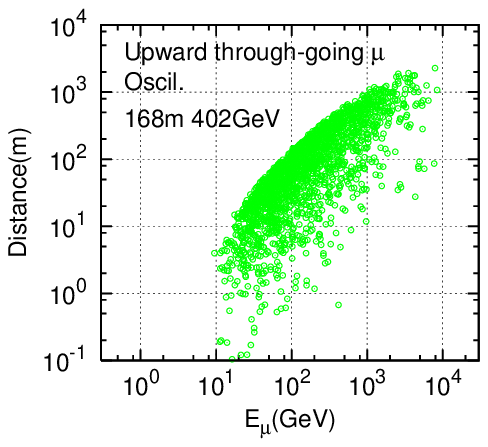}
\caption{}
\label{fig:14}
\end{minipage}\hspace{0.5pc}%
\begin{minipage}[t]{12pc}
\centering
\includegraphics[width=12pc]{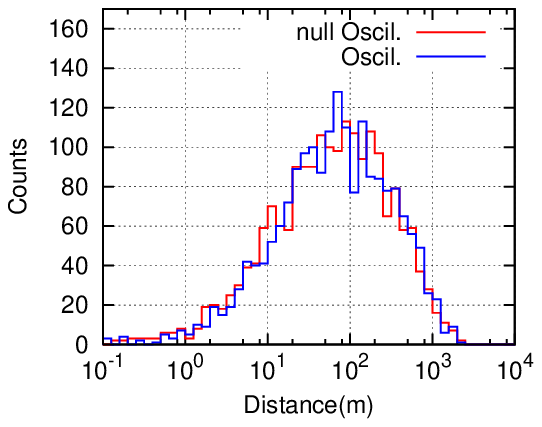}
\caption{}
\label{fig:15}
\end{minipage}
\end{figure}

In Figure \ref{fig:13}, we give the correlation diagram between the energies of the emitted muons and the distance from their occurrence points up to the detector for \textit{Upward Through Going Muon Events} for null oscillation, pursuing the muons concerned in stochastic manner. The average energy of the emitted muons is 390 GeV and their average distance 161 meter.

In Figure \ref{fig:14}, we give the corresponding quantities for oscillation. The average energy of the emitted muons is 402 GeV and their average distance is 168 meters. It is easily understood from the Figure \ref{fig:13} and Figure \ref{fig:14} that the contribution from higher energy part of the muon could not negligible, which is larger beyond the expectation by SK whose analysis is based on the average theory for muon propagation. Furthermore, comparing the Figure \ref{fig:13} with Figure \ref{fig:14}, we could not find clear differences between them, which denote the difficulty in extracting the SK oscillation parameters from the analysis of the \textit{Upward through Going Muon events}.

\vspace{-2mm}

\vspace{-3mm}
\begin{minipage}[t]{1pc}
\end{minipage}\hspace{2pc}%
\begin{minipage}[t]{33pc}
\noindent
{\small \textbf{Figure \ref{fig:13}. }Correlation diagram between the emitted muon energies and their distance from their generation points to the detector for null oscillation in the case of \textit{Upward Through Going Muon Events}.
}\\
\noindent
{\small \textbf{Figure \ref{fig:14}. }Correlation diagram between the emitted muon energies and their distance from their generation points to the detector for the oscillation in the case of \textit{Upward Through Going Muon Events}.
}\\
\noindent
{\small \textbf{Figure \ref{fig:15}. }The distance distribution for \textit{Upward Through Going Muon Events} in both cases of null oscillation and the oscillation.
}
\end{minipage}

\vspace{4mm}
\noindent
In Figure \ref{fig:15}, we give the distance distribution for \textit{Upward Through Going Muon Events} in the cases of both null oscillation and the oscillation for 1645.9 SK live days. The distance denotes the interval from the generation point of the events concerned to the detector.  It is easily understood from the figure that we could not discriminate the difference in both cases well, which suggests difficulty in extracting the definite SK neutrino oscillation parameters.

\vspace{-3mm}
\begin{figure}[h]
\begin{minipage}[h]{18pc}
\centering
\includegraphics[width=18pc]{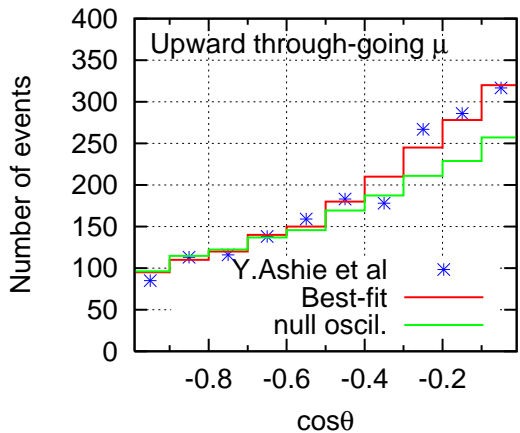}
\end{minipage}\hspace{1pc}%
\begin{minipage}[h]{18pc}
\caption{Comparison of the zenith angle distribution for \textit{Upward Through Going Muon Event} by SK with their Monte Carlo Results }
\label{fig:16}
\end{minipage}
\end{figure}%
\vspace{-3mm}
\noindent
Now, we return the SK original data for \textit{Upward Through Going Muon Events} which we reproduce it in Figure \ref{fig:16}. Now, we point out clear contradiction in Figure \ref{fig:16}. The first to be examined  is that frequency of the experimental data is significantly larger than that of their Monte Carlo Simulation without oscillation in the wider region (-0.6, 0.0) for $\cos \theta$ and the second is rather nice agreement between them in the region (-1.0,-0.6) for $\cos \theta$ . The first denotes that usually we should expect smaller frequency with oscillation than that without oscillation. However, here, the situation becomes reverse which contradict with the commonsense.  Furthermore, in the region (1.0,-0.6) for $\cos \theta$, the effect of the neutrino oscillation should be pronounced due to longer path length of the incident neutrino. However, the real situation seems to indicate no oscillation.
In this case, there are three possible interpretations to such a situation. Namely, [a] Monte Carlo Simulation is not unreliable, while the experimental data is reliable including statistical uncertainty, [b] Monte Carlo Simulation is reliable, while statistics of the experimental data is too small to give larger frequency of the neutrino events compared with Monte Carlo events, [c] Neither experimental data nor Monte Carlo data is unreliable to exclude the definite oscillation parameters.

Furthermore, from the view point of the comparison between experimental data and Monte Carlo one, we should make two comments: the fist is that SK Monte Carlo simulation does not consider the real behaviors of the muons in stochastic manner, but their average behaviors only. The second is that it is not suitable way to compare experimental data for 1645.9 days with Monte Carlo one for 100 live years, because of big difference in statistics between them. The normalization of 100 live yeas to 1645.9 days never guarantees the validity of their transformation without examination on the validity concerned.

To clarify the contradiction in Figure \ref{fig:16}, it is necessary for SK to disclose their Monte Carlo procedure in detail. 

\subsection{Zenith angle distribution for the Stopping Muon Events}
\vspace{-5mm}
\begin{figure}[h]
\begin{minipage}[t]{18pc}
\centering
\includegraphics[width=18pc]{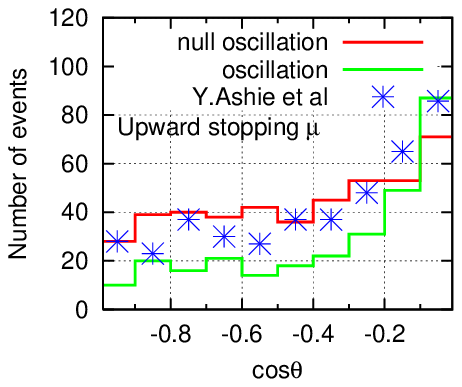}
\caption{Zenith angle distribution for \textit{Upward Stopping Muon Events} for 1645.9 live-days}
\label{fig:17}
\end{minipage} \hspace{1pc}%
\begin{minipage}[t]{18pc}
\centering
\includegraphics[width=18pc]{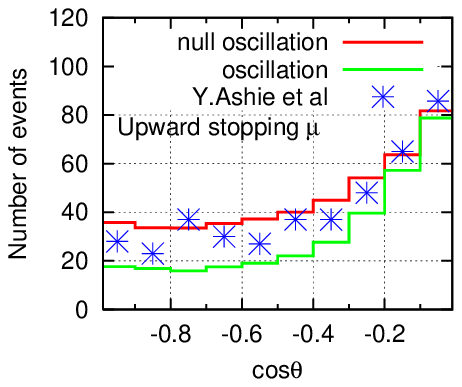}
\caption{Zenith angle distribution for \textit{Upward Stopping Muon Events} for 164590 live-days, which is normalized to 1645.9 live-days.}
\label{fig:18}
\end{minipage}
\end{figure}%
\vspace{-3mm}
\noindent
It is natural to think that \textit{Upward Stopping Muon Events} are much influenced by fluctuation, compared with \textit{Upward Through Going Muon Events}, because they are more directly influenced by the complicated geometry of the detector with regard to the incident neutrinos than \textit{Upward Through Going Muon Events} are done.

In Figure \ref{fig:17}, we compare the SK experimental data for 1645.9 live days with our data with oscillation and without oscillation for \textit{Upward Stopping Muon Events} for the same 1645.9 live days. From the figure, also, we could conclude that SK experimental data neither agree with our data with oscillation nor that without oscillation. In the figure, it should be noticed that the number of the neutrino events with oscillation is larger than that without oscillation due to fluctuation effect for $\cos \theta = -0.05 $ $(0.0 \sim -0.10 )$ . 

In Figure \ref{fig:18}, we give our results with oscillation and without oscillation for 164590 live days together with SK experimental data which, in principle, should not be allowed to be compared with our data due to big difference in statistics as stated in the case of \textit{Upward Through Going Muon Events} (see, also footnote on page 13). Compared Figure \ref{fig:18} with Figure \ref{fig:17} we understand that the difference between quantity with oscillation and that without oscillation in Figure \ref{fig:18} decrease compared with that in Figure \ref{fig:17} and both histograms in Figure \ref{fig:18} become smooth compared with those in Figure \ref{fig:17}, which is evidence that "average value " attain at ideal average value.

\vspace{-4mm}
\begin{figure}[h]
\begin{minipage}[t]{12pc}
\centering
\includegraphics[width=12pc]{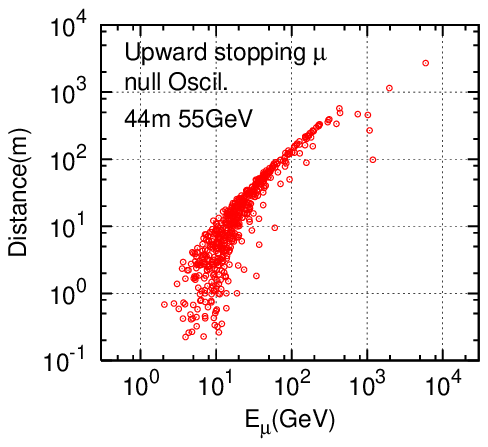}
\caption{\label{fig:19} }
\end{minipage}\hspace{0.5pc}%
\begin{minipage}[t]{12pc}
\centering
\includegraphics[width=12pc]{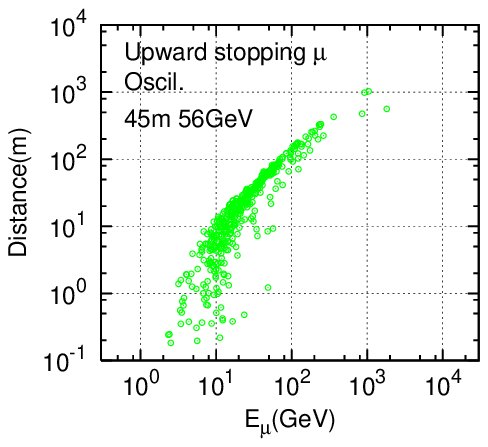}
\caption{\label{fig:20} }
\end{minipage}\hspace{0.5pc}%
\begin{minipage}[t]{12pc}
\centering
\includegraphics[width=12pc]{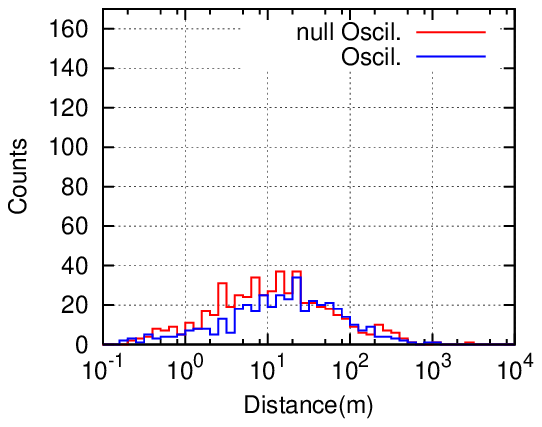}
\caption{\label{fig:21} }
\end{minipage}
\end{figure}%

\vspace{-4mm}
\begin{minipage}[t]{1pc}
\end{minipage}\hspace{2pc}%
\begin{minipage}[t]{33pc}
\noindent
{\small \textbf{Figure \ref{fig:19}. }Correlation diagram between the emitted muon energies and their distance from their generation points to the detector for null oscillation in the case of \textit{Upward Stopping Muon Events}.} \\
\noindent
{\small \textbf{Figure \ref{fig:20}. }Correlation diagram between the emitted muon energies and their distance from their generation points to the detector for the oscillation in the case of \textit{Upward Stopping Muon Events}.} \\
\noindent
{\small \textbf{Figure \ref{fig:21}. }The distance distribution for \textit{Upward Stopping Muon Events} in both cases of null oscillation and the oscillation.}
\end{minipage}

\vspace{4mm}
\noindent
In Figure \ref{fig:19}, we give the correlation diagram between the emitted muon energies and their distances in \textit{Stopping Muon Events} for null oscillation. The average energy of the emitted muons concerned is 55 GeV and their average distance is 44 meters. In Figure \ref{fig:20}, we give the similar diagram for SK oscillation parameters as Figure \ref{fig:19}. The average energy of the muons is 56 GeV and their average distance is 45 meters. The situation around Figure \ref{fig:19} and Figure \ref{fig:20} is essentially same as in that around Figure \ref{fig:13} and \ref{fig:14}.

In Figure \ref{fig:21}, we give the distance distribution for \textit{Upward Stopping Muon Events} in both cases of null oscillation and the oscillation. It is reasonable to say from the figure that we could not find the significant difference in the both distance distribution, which may lead to the difficulty in extracting the definite oscillation parameters like SK do.

\subsection{The energy spectrum for the parent neutrinos and their daughter muons in the presence of the neutrino oscillation}
\vspace{-5mm}
\begin{figure}[h]
\begin{minipage}[t]{12pc}
\centering
\includegraphics[width=10pc]{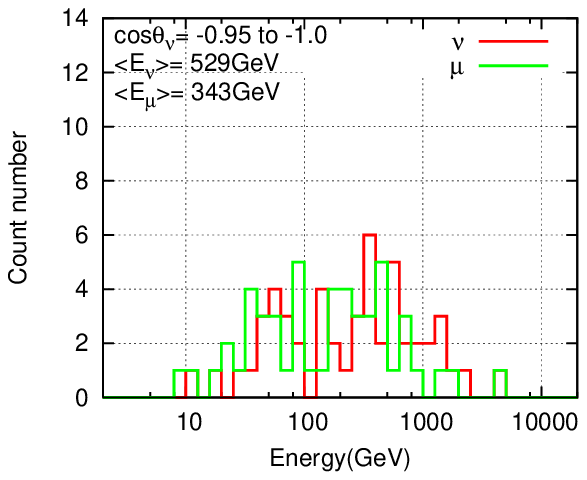}
\caption{\label{fig:22} }
\end{minipage}\hspace{0.5pc}%
\begin{minipage}[t]{12pc}
\centering
\includegraphics[width=10pc]{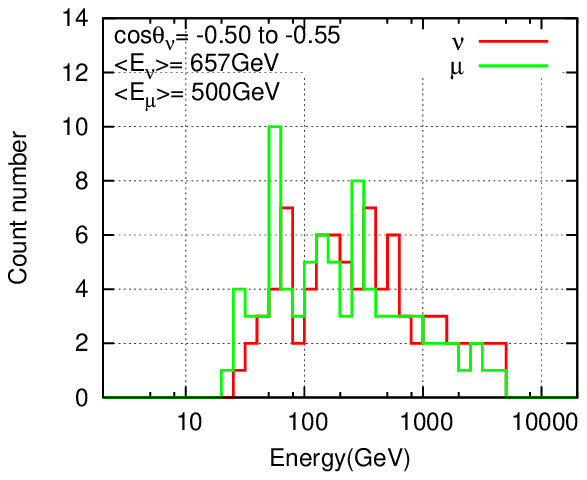}
\caption{\label{fig:23} }
\end{minipage}\hspace{0.5pc}%
\begin{minipage}[t]{12pc}
\centering
\includegraphics[width=10pc]{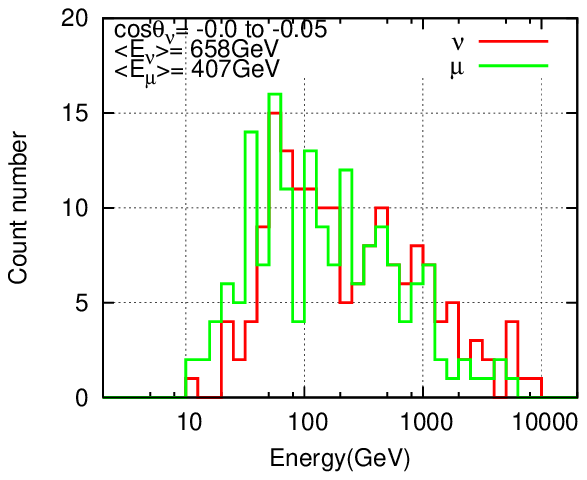}
\caption{\label{fig:24} }
\end{minipage}
\end{figure}%

\vspace{-4mm}
\begin{minipage}[t]{1pc}
\end{minipage}\hspace{2pc}%
\begin{minipage}[t]{33pc}
\noindent
{\small \textbf{Figure \ref{fig:22}. }Energy spectra for parent neutrinos and daughter muons for \textit{Upward Through Going Muon Events} for $\cos \theta _{\nu }= -0.957$ $( -0.95 \sim -1.0 )$.} \\
\noindent
{\small \textbf{Figure \ref{fig:23}. }Energy spectra for parent neutrinos and daughter muons for \textit{Upward Through Going Muon Events} for $\cos \theta _{\nu }= -0.525$ $( -0.500 \sim -0.550 )$.} \\
\noindent
{\small \textbf{Figure \ref{fig:24}. }Energy spectra for parent neutrinos and daughter muons for \textit{Upward Through Going Muon Events} for $\cos \theta _{\nu }= -0.025$ $( -0.0 \sim -0.05 )$.} 
\end{minipage}

\vspace{0mm}
\begin{figure}[h]
\begin{minipage}[t]{12pc}
\centering
\includegraphics[width=10pc]{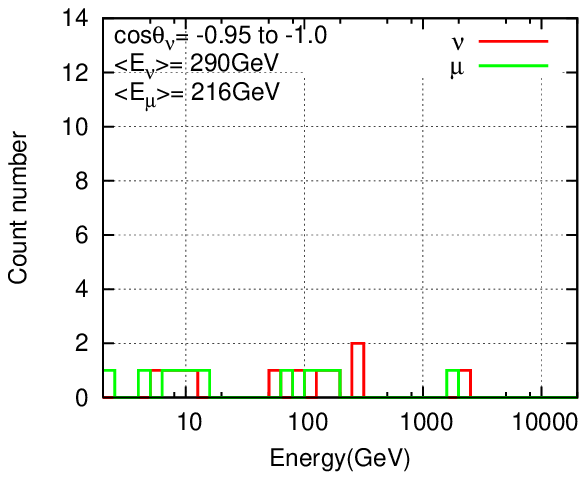}
\caption{\label{fig:25} }
\end{minipage}\hspace{0.5pc}%
\begin{minipage}[t]{12pc}
\centering
\includegraphics[width=10pc]{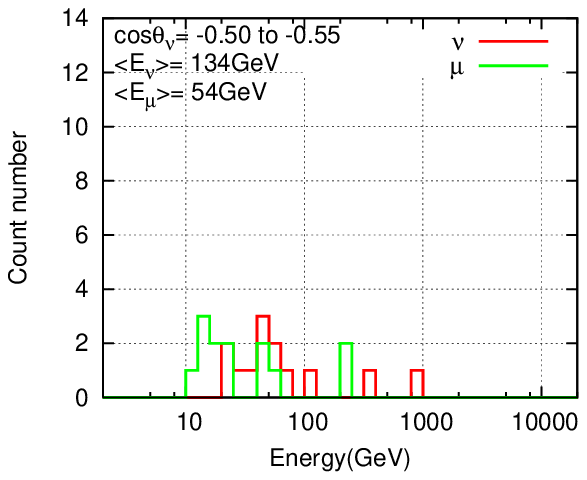}
\caption{\label{fig:26} }
\end{minipage}\hspace{0.5pc}%
\begin{minipage}[t]{12pc}
\centering
\includegraphics[width=10pc]{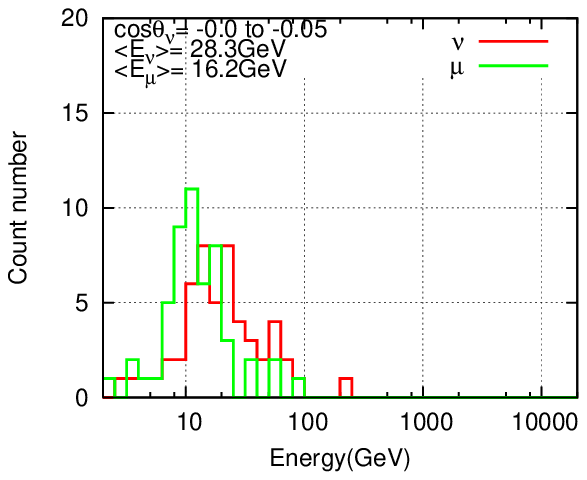}
\caption{\label{fig:27} }
\end{minipage}
\end{figure}%

\vspace{-2mm}
\begin{minipage}[t]{1pc}
\end{minipage}\hspace{2pc}%
\begin{minipage}[t]{33pc}
\noindent
{\small \textbf{Figure \ref{fig:25}. }Energy spectra for parent neutrinos and daughter muons for \textit{Upward Stopping Muon Events} for $\cos \theta _{\nu }= -0.957$ $( -0.95 \sim -1.0 )$.} \\
\noindent
{\small \textbf{Figure \ref{fig:26}. }Energy spectra for parent neutrinos and daughter muons for \textit{Upward Stopping Muon Events} for $\cos \theta _{\nu }= -0.525$ $( -0.500 \sim -0.550 )$.} \\
\noindent
{\small \textbf{Figure \ref{fig:27}. }Energy spectra for parent neutrinos and daughter muons for \textit{Upward Stopping Muon Events} for $\cos \theta _{\nu }= -0.025$ $( -0.0 \sim -0.05 )$.}
\end{minipage}

\vspace{4mm}
\noindent
Now, we examine energy spectra of the parent neutrinos and their daughter muons for the \textit{Upward Through Going Muon Events} and the \textit{Stopping Muon Events} with different $\cos \theta _{\nu }(\theta _{\mu })$.

In Figures  \ref{fig:22},  \ref{fig:23} and \ref{fig:24}, we give energy spectra of \textit{Upward Through Going Muon Events} for $\cos \theta _{\nu }= -0.975$, (upward) $-0.525$(diagonally) and $-0.025$(horizontally), respectively for SK live days, 1645.9 days.
In Figure \ref{fig:25}, \ref{fig:26} and \ref{fig:27}, we give the energy spectra of the \textit{Stopping Muon Events} for the same   for the same SK live days.  It seems to be clear from Figures \ref{fig:22} to \ref{fig:27} that all these energy spectra are never smooth which denote insufficient statistics to extract definite conclusions, particularly as shown in  Figure \ref{fig:24}.

In Figures \ref{fig:22}, \ref{fig:23} and \ref{fig:24}, we give the average energies for parent neutrinos and daughter muons ( in parentheses) in the case of the \textit{Upward Through Going Muon Events} for $\cos \theta _{\nu }= -0.975$, $-0.525$ and $-0.025$, respectively, namely 529GeV (343GeV), 657GeV (500GeV) and 658 GeV (407GeV).  The corresponding average energy for the parent neutrino in SK is $\sim $100 GeV~\cite{Oyama.Y.1989_HaSaNi.PhDthesis} which is pretty smaller than that our values. In Figure \ref{fig:25}, \ref{fig:26} and \ref{fig:27}, we give the average energies of the \textit{Stopping Muon Events} for the same $\cos \theta _{\nu }$, 290GeV (216GeV), 134GeV (54GeV) and 28GeV (16GeV), respectively. The corresponding average energy for neutrino  by SK is $\sim $ 5 GeV~\cite{Oyama.Y.1989_HaSaNi.PhDthesis} which is also too  smaller than our corresponding values.

As for obtaining the average energy of the \textit{Upward Through Going Muon Events}, SK utilize the average theory which is unsuitable, as already mentioned, for the analysis of real behavior of high energy muon and, therefore, they could not consider the contribution from higher energy muons correctly.

As for the estimation of the average energy of \textit{Stopping Muon events}, it seems to be curious that their average energy by SK is nearly same as or a little higher than that of Multi-GeV Events. Because Multi-GeV Events comes from essentially single meson scattering, while \textit{Stopping Muon Events} come from deep inelastic scattering, and the effective energy for single meson scattering is too lower than that of deep inelastic scattering (see, Figure \ref{fig:15} in the SK paper~\cite{Ashie.Y.2005}). The difference in the effective energies between \textit{Upward through Going Muon Events} and \textit{Stopping Muon Events} comes from the difference in the effective ranges from the same elementary process, say, deep inelastic scattering.  Also, SK estimate that generation points up to the detector distributed over 0 meters to several hundred meters~\cite{Oyama.Y.1989_HaSaNi.PhDthesis}. Too small estimation on average energy as well as the generation points for \textit{Upward Stopping Muon} by SK  might come from the utilization of the average theory on energy loss of the muon.

Here, let us remark the relation between oscillatory nature of the primary neutrino interaction energy spectrum and their influence over \textit{Upward Through Going Muon Events} and \textit{Stopping Muon Events}.  It s easily understood from Figures \ref{fig:22} to \ref{fig:27}, referring to Figure \ref{fig:07}, that neutrinos contributed to both \textit{Upward Through Going Muon Events} and \textit{Stopping Muon Events }for all their zenith angles come from pretty high energies which belong non-oscillatory part of the incident neutrino interaction energy spectrum under the specified neutrino oscillation parameters by SK.

In other word, we should say that the analysis of \textit{Upward Through Going Muon Events} and \textit{Stopping Muon Events} are not suitable means for proving the existence of neutrino oscillation under SK neutrino oscillation parameters.

\subsection{Comparison of Monte Carlo Simulation in enough statistics by SK with the corresponding one by our computer Numerical Experiment}

Here, we compare SK Monte Carlo Simulation in the case of null oscillation for enough statistics with our corresponding Computer Numerical Experiment.

Now, we compare SK Monte Carlo results for 100 live years with ours for 450 live years for both \textit{Upward Through Going Muon Events} and \textit{Stopping Muon Events} in the case of null oscillation.  Strictly speaking, SK results for 100 live years should be compared with ours for same live year. Or our results for 450 live years should be compared with the SK results for same live years. However, as SK data for 100 live years are supposed to attain at the enough statistics which is free from fluctuation, compared with that for 1645.9 days, we might be allowed to compare directly SK data for 100 live years with our data for 450 live years.

\vspace{-3mm}
\begin{figure}[h]
\begin{minipage}[t]{18pc}
\includegraphics[width=18pc]{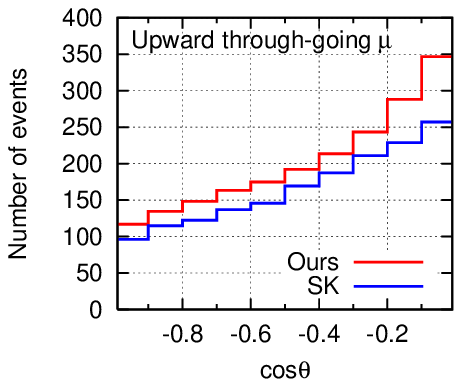}
\caption{Our Monte Carlo result and SK Monte Carlo ones for null oscillation for \textit{Upward Through Going Muon Events}. Both Monte Carlo results are normalized to 1645.9 live-days.}
\label{fig:28}
\end{minipage} \hspace{1pc}%
\begin{minipage}[t]{18pc}
\includegraphics[width=18pc]{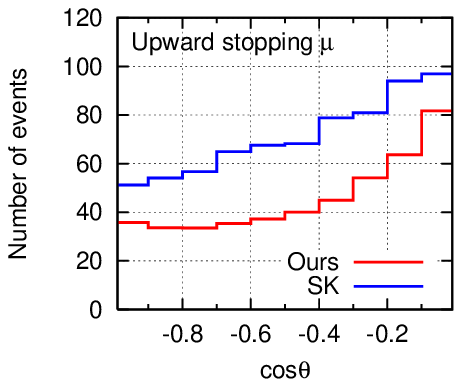}
\caption{Our Monte Carlo results and SK Monte Carlo ones for null oscillation for \textit{Upward Stopping Muon Events}. Both Monte Carlo results are normalized to 1645.9 live-days.}
\label{fig:29}
\end{minipage}
\end{figure}%
\vspace{-3mm}

From Figure \ref{fig:28} we understand that general tendency is similar to each other and our results give larger value than SK results. It is reasonable, if we take into account of the fact that SK neglect the range fluctuation in high energy muon.

In Figure \ref{fig:29}, we give the direct comparison between SK results and our results for \textit{Upward Stopping Muon Events}. Here, SK results give larger values than our results, on the contrary to Figure \ref{fig:28}. This reversed situation between Figure \ref{fig:28} and Figure \ref{fig:29} seems to be curious, if we take into account of the range fluctuation of high energy muon. Also, tendencies in the SK histograms in Figure \ref{fig:29} are somewhat unnatural in lack in smoothness in spite of huge statistics.
 
It is desirable for SK to disclose their essential information on their Monte Carlo method for the case of the null oscillation in order to clarify the difference between SK's and ours. The techniques for our Monte Carlo method fully disclose in the present paper. As for our Monte Carl Simulation for the neutrino events inside the detector, we disclose our procedure in detail in the subsequent papers.

\section{Conclusion}
Thus, the followings should be concluded: \\
\noindent
(a) The fluctuation effect in the high energy muon behavior from the neutrino event (due to deep inelastic scattering) primarily comes from the uniformly randomness of the  generation point of the neutrino events concerned. This effect could not be treated in the correct manner in the Detector Simulation by SK.\\
\noindent
(b) 
Also, the consideration of the range fluctuation in high energy muon is inevitable, although this effect is pretty smaller than that from the generation point of the neutrino events. We consider this effect exactly in the stochastic manner. On the other hand, SK analyzes the behavior of muons in the average theory, namely, in the concept of the average energy loss and average range, not treating  them in stochastic manner. Combined of [a] with [b], we claim that experimental results for the \textit{Upward Through Going Muon Events} and the \textit{Stopping Muon Events} do not show the evidence under SK neutrino oscillation parameters.\\
\noindent
(c) Generally speaking, it is difficult  to imagine rationally to extract so definite conclusion around neutrino oscillation parameters, say, $\sin^{2}2\theta $ and $\Delta m^{2}$ through the analysis of the \textit{Upward through Going Muon Events} and \textit{Stopping Muon Events}. Because, there are too few information around such categorized events to extract the definite conclusion on the neutrino oscillation parameters. Such categorized events provide the direction of the incident neutrino only. \textit{Upward Through Going Muon Events} never provide us the energies of the incident neutrinos, their interaction point and the energies of the emitted muons, their stopping points, giving the information on their penetration through the detector only. \textit{Stopping Muon Events} gives the information that they stop inside the detector and other information not obtained by them are same as the \textit{Upward Through Going Muon Events}. \\
\noindent
(d) Even if the SK neutrino oscillation parameters exist really, it is difficult to estimate them from the analysis of \textit{Upward through Going Muon Events} and \textit{Stopping Muon Events}. Because, the effective energies to detect the SK neutrino oscillation parameters, say, smaller than 40 GeV,  do not lie in the energy region which contribute to the production of \textit{Upward Through Going Muon Events} and the \textit{Stopping Muon Events}.\\
\noindent
(e) By the definition of the \textit{Upward Through Going Muon Events} and the \textit{Stopping Muon Events} in the SK, the origins of both events are same(deep inelastic scattering)and difference between them is that the former average range is longer than that of the latter so that the former average energy is higher than that of the latter. However, the latter average is so low which is rather near to the average energy due to multi-GeV energy events. This is difficult to be understood. Also, the zenith angle distribution for \textit{Upward through Going Muon events} is difficult to be understood in comparison with their Monte Carlo simulation with 100 live years.\\
\noindent
(f) Even if we could obtain the SK neutrino oscillation parameters from the analysis of both \textit{Upward through Going Muon Events} and the \textit{Stopping Muon Events}, present statistics for experimental data obtained by 1645.9 live days is not enough for getting the SK neutrino oscillation parameters.\\
\noindent
(g) As the effective energy to detect oscillatory signature of the neutrino oscillation parameters lies below 40 GeV under the assumption of the SK neutrino oscillation parameters, first of all, one should concentrate to analyze the single ring muon event among the Fully Contained Events, the most clear cut event from the quasi elastic scattering in GeV energy region.

Generally speaking, it is rather difficult to extract so definite physical quantities, such as, neutrino oscillation parameters from the atmospheric neutrino beams which include much uncertainty factors, such as, the incident neutrino energy spectrum. The throughout and concentrated analysis of the single ring lepton events are inevitable, if we really extract definite parameters from the cosmic ray experiment in spite of the existence of much uncertainty factors involved.

We are now going to complete our works along this line and will present our results  from the computer numerical experiment, focusing on the analysis of single-ring muon events as a series of subsequent papers elsewhere.


\end{document}